\documentclass{article} 
\usepackage[T1]{fontenc}
\usepackage{latex8}
\usepackage{charter}
\usepackage{mathrsfs}
\usepackage{amsfonts}
\usepackage[lined,algonl,ruled]{algorithm2e}
\usepackage{tikz}
\usepackage{gantt}
\usetikzlibrary{arrows}

\newtheorem{Def}{Definition}
\newtheorem{Prop}{Property}

\usepackage[sumlimits]{amsmath}
\usepackage{xargs}

\newcommand{\drop}[1]{}

\newcommand{\TD}{{\mathscr{D}}}
\newcommand{\UB}{\textsf{UB}}
\newcommand{\HV}{\textsf{HV}}

\begin{document}

\title{Partitioned scheduling of multimode multiprocessor\\real-time systems with temporal isolation}

\author{Joël Goossens
\and
Pascal Richard
}

\maketitle
\thispagestyle{empty}

\begin{abstract}
We consider the partitioned scheduling problem of multimode real-time systems upon identical multiprocessor platforms. During the execution of a multimode system, the system can change from one mode to another such that the current task set is replaced with a new one. In this paper, we consider a synchronous transition protocol in order to take into account mode-independent tasks, i.e., tasks of which the execution pattern must not be jeopardized by the mode changes. We propose two methods for handling mode changes in partitioned scheduling. The first method is offline/optimal and computes a static allocation of tasks schedulable and respecting both tasks and transition deadlines (if any). The second approach is subject to a sufficient condition in order to ensure online First Fit based allocation to satisfy the timing constraints.
\end{abstract}

%------------------------------------------------------------------------- 
\Section{Introduction}
Hard real-time systems require both functionally correct executions and \emph{results that are produced on time}. Currently, numerous techniques exist that enable engineers to design real-time systems while guaranteeing that all their temporal requirements are met. These techniques generally model each functionality of the system by a \emph{recurrent} task, characterized by a computing requirement and an activation rate. Commonly, real-time systems are modeled as a set of such tasks. However, some applications exhibit multiple behaviors issued from several operating modes (e.g.,~an initialization mode, an emergency mode, a fault recovery mode, etc.), where each mode is characterized by its own set of functionalities, i.e., its set of tasks. During the execution of such \emph{multimode} real-time systems, switching from the current mode (called the \emph{old-mode}) to another one (the \emph{new-mode} hereafter) requires to substitute the current executing task set with the set of tasks of the new-mode. There are tasks however --- called \emph{mode-independent tasks} in the literature --- which should execute in \emph{every} mode and such that their activation pattern must not be jeopardized during the transition between those modes\footnote{In practice, mode-independent tasks typically allow to model daemon functionalities and low-level control loops.}. 

Transition scheduling protocols are often classified with respect to the way they schedule the old- and new-mode tasks during the transitions. In the literature (see for instance~\cite{JoAlfons:04}), the following definitions are used.

\begin{Def}[Synchronous/Asynchronous protocol] 
\label{def:Multimode:synchronous_asynchronous}
A mode-change protocol is said to be synchronous if it schedules new-mode tasks only when \emph{all} the old-mode tasks have completed. Otherwise, it is said to be asynchronous. 
\end{Def}

\begin{Def}[Protocol with/without periodicity]
\label{def:Multimode:with_without_periodicity}
A mode-change protocol is said to be ``with periodicity'' if and only if it is able to deal with mode-independent tasks. Otherwise, it is said to be ``without periodicity''.
\end{Def}

In this research we consider identical multiprocessor platforms and \emph{partitioned} scheduling, i.e.\@ the tasks are \emph{statistically} assigned to a processor of the platform, at runtime task/job migration is forbidden. Our motivation of this framework is driven by applications which require a \emph{temporal isolation} between the various partitions/components. Such requirement are common for \emph{certified} avionic or automotive applications (e.g. DO-178B, DO-254 and ISO 26262 standards). Additional motivation of our framework in the technologies like Ada 2012 and integrated tool-chain~\cite{saez2012integrated} which enable the design and implementation of partitioned multimode multiprocessor applications.

\SubSection{Related work}
Numerous scheduling protocols have already been proposed in the \emph{uni}processor case to ensure the transition between modes (see~\cite{JoAlfons:04} for a survey of the literature about this uniprocessor problem). For multiprocessor platforms most of the contributions concern \emph{global} scheduling approaches where job migrations are allowed (see e.g.\@ the works of Nélis and colleagues~\cite{multimode:09,MeumeuNelisGoossens:10,Nelis:10,nelis2011global}).

Concerning the \emph{partitioned} approach we can report~\cite{marinho2011partitioned} the short paper by Marinho et al.\@ which formalizes the scheduling problem and shows two counter-intuitive phenomenons. Emberson and Bate~\cite{emberson2007minimising} proposed heuristics to handle the mode change; unfortunately they do not provide any timing guarantee to bound the transition delays.    

\paragraph{This research.} In this work we consider the scheduling of sporadic implicit-deadline real-time tasks upon identical multiprocessor platforms. Each mode is characterized by a static task partitioning, each processor has its own local and optimal scheduler (EDF typically). We consider synchronous mode change protocol and  mode-independent tasks which cannot migrate across the platform. Our contributions are twofold. We propose two methods for handling mode changes. The first method is offline/optimal and computes a static allocation of tasks schedulable and respecting both tasks and transition deadlines (if any). The second approach is subject to a sufficient condition in order to ensure \emph{online} First-Fit based allocation will satisfy the timing constraints.

%------------------------------------------------------------------------- 
\Section{Protocol definition and properties}

\SubSection{Software and hardware architecture}

The software architecture is defined by a set of  sporadic tasks with implicit-deadlines. Each task $\tau_i$, $1 \leq i \leq n$ is  defined by a worst-case execution time $C_i$, a minimal interval between two successive release $T_i$. Thus, $\tau_i$ generates an infinite collection of jobs and every job must be completed before the earliest next job release (i.e., implicit-deadline). The task utilization $U_i=\frac{C_i}{T_i}, 1 \leq i \leq n$ is the fraction of time required by the task execution (upon a single processor). The platform utilization is defined by $U_{\text{sum}}= \sum_{i=1}^n U_i$ and the maximum task utilization is $U_{\max}=\max_{i=1}^n U_i$. 

The multiprocessor platform is defined by $m$ identical processors ($\pi_{1},\ldots,\pi_{m}$) with a shared memory. We assume that task allocations and preemptions are performed without any incurring cost. In practice, the corresponding delays are taken  into account in the timing analysis of tasks (i.e., definition of worst-case execution times).

The different transitions between modes in the real-time software can be described by an directed graph. Vertices are the modes and directed edges models are the transitions between a source mode to a destination mode. Edges are labeled by the worst-case transition delay $L_i$ between the two connected modes. According to the protocol that will be consider in this paper, the transition delay will only depend of the source mode whatever the destination mode (this will be discussed in details in the next section). That is the reason why labels $L_i$ on every edge are only indexed according to the source mode.  

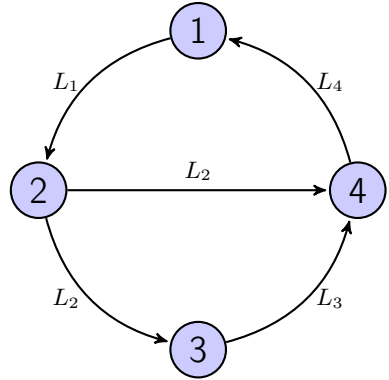
\begin{figure}
\begin{center}
\begin{tikzpicture}[->,>=stealth',shorten >=1pt,auto,node distance=3cm,
  thick,main node/.style={circle,fill=blue!20,draw,font=\sffamily\Large\bfseries}]

  \node[main node] (1) {1};
  \node[main node] (2) [below left of=1] {2};
  \node[main node] (3) [below right of=2] {3};
  \node[main node] (4) [below right of=1] {4};

  \path[every node/.style={font=\sffamily\small}]
    (1) edge [bend right] node[left] {$L_1$} (2)
    (2) edge node {$L_2$} (4)
        edge [bend right] node[left] {$L_2$} (3)
    (3) edge [bend right] node[right] {$L_3$} (4)
    (4) edge [bend right] node[right] {$L_4$} (1);
\end{tikzpicture}
\end{center}
\caption{\label{fig:graph} Mode change transition graph example}
\end{figure}

Usually, all possible transitions between modes can be defined at the design stage and thus such a graph is a priori known. We make no particular assumption on this graph that models all possible transition modes during the system life. Figure~\ref{fig:graph} presents a mode change graph example with 4 modes and 5 possibles transitions between them. Notice that from Mode~2, two possible mode changes are possible to reach either Mode~3 or Mode~4. This selected destination mode can thus depends on the state of the real-time system when the mode change request occurs. 

\SubSection{Protocol definition and objective}
The tasks are executed according to the system modes. Transitions between modes are initiated by a specific event called a \emph{mode change request} (MCR). Mode independent tasks (MI tasks) are executed in \emph{every} mode whereas mode dependent tasks (MD tasks) that are executed in the mode in which they belong to. A mode dependent task $\tau_i$ is subjected to a transition deadline $\TD_i$ that ensures that the first job of $\tau_i$ must be completed after $\TD_i$ time units after a MCR initiating the mode in which $\tau_i$ belongs to. Such a deadline is used to take into account the mode transition latency (i.e., the delay to stop MD jobs after a mode change request). After a mode change request at time $a$, the first released job of the MD task $\tau_i$ must complete occur before time instant: $r_i \leq a+\TD_i$ (or equivalently must be released before time instant: $r_i \leq a+\TD_i - T_{i}$).

We consider synchronous mode-change protocol~\cite{JoAlfons:04}. This means that  whenever a MCR occurs, MD tasks in the running mode are no longer recurrent and the mode change is completed when all jobs of MD tasks are completed. At this time, the MD tasks of the new mode are then \emph{enabled} simultaneously, i.e., those tasks can generate jobs (sporadically). 

\begin{figure*}
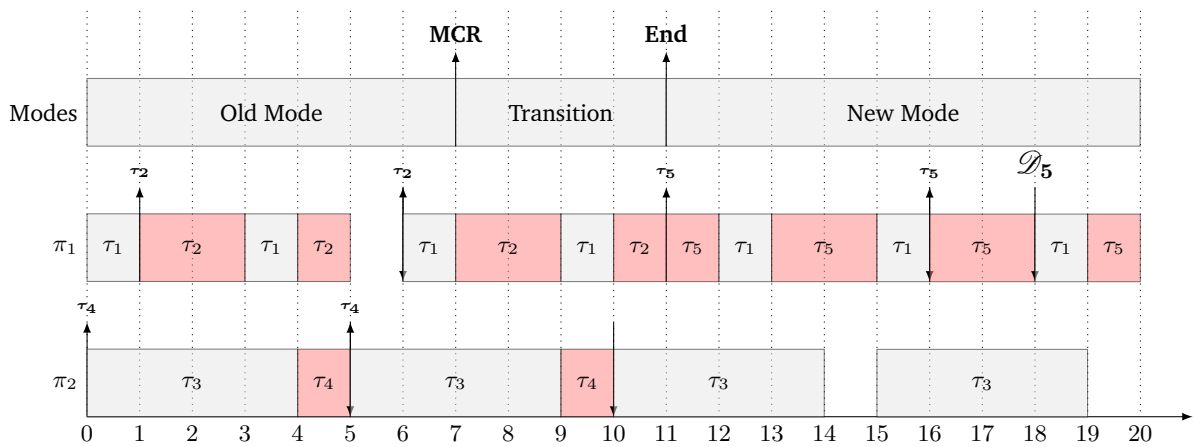

\begin{center}
\begin{gantt}{20}{3}{0.7}{1.8}
\activityrow{Modes} \activity{Old Mode}{7} \request{\small{\textbf{MCR}}} \activity{Transition}{4} \request{\small{\textbf{End}}} \activity{New Mode}{9} 
 
%\activityrow{}
 
\activityrow{$\pi_1$}\activity{$\tau_{1}$}{1} \request{$\boldsymbol{\tau_2}$} \lateactivity{$\tau_2$}{2} \activity{$\tau_{1}$}{1}  \lateactivity{$\tau_2$}{1} \lagtime{}{1}  \request{$\boldsymbol{\tau_2}$}\deadline{}\activity{$\tau_{1}$}{1}  \lateactivity{$\tau_2$}{2} \activity{$\tau_{1}$}{1}  \lateactivity{$\tau_2$}{1} \request{$\boldsymbol{\tau_5}$} \lateactivity{$\tau_5$}{1}  \activity{$\tau_{1}$}{1}\lateactivity{$\tau_5$}{2}  \activity{$\tau_{1}$}{1}\deadline{}\request{$\boldsymbol{\tau_5}$}  \lateactivity{$\tau_5$}{2} \deadline{\large{$\boldsymbol{\TD_5}$}}\activity{$\tau_{1}$}{1}  \lateactivity{$\tau_5$}{1} \lagtime{}{1}

%\activityrow{}

\activityrow{$\pi_2$}  \request{$\boldsymbol{\tau_4}$}\activity{$\tau_{3}$}{4}\lateactivity{$\tau_4$}{1}\deadline{} \request{$\boldsymbol{\tau_4}$} \activity{$\tau_3$}{4}\lateactivity{$\tau_{4}$}{1} \deadline{}\activity{$\tau_{3}$}{4} \lagtime{}{1} \activity{$\tau_{3}$}{4}

%\activityrow{$\tau_4$ (new)} \lagtime{}{13}\request{} \activity{$\tau_4$}{2} \lagtime{}{1}  \lagtime{}{1} \deadline{\large{$\boldsymbol{\TD_4}$}} \request{} \activity{$\tau_4$}{2} 

\end{gantt}
\end{center}
\caption{\label{fig:deadline} Mode change request with a synchronous mode change protocol upon 2 processors}
\end{figure*}

The system is characterized by a set $I$ of MI tasks that will still running continuously after every MCR. After a mode change request (MCR), every MD task is disabled. New tasks will be enabled when all old MD tasks will be completed. Thus, the task system and the synchronous mode change protocol leads to the following assumptions:
\begin{itemize}
\item every MI task is running in every mode and its processor allocation is known a priori and will never change at runtime.
\item none of the MD tasks belongs to several modes.
\end{itemize}

We think that these assumptions correspond to many real-world applications. They allow to select which mode is activated according the environment whenever a MCR arises. Furthermore, as we shall see, it simplifies the definition of the mode transition change latency as it will be presented in Property~\ref{prop:delay}. 

In order to illustrate the synchronous protocol behavior, we present a simple example. Consider the tasks presented in Table~\ref{tab:mode}. Every task is assumed here to be allocated to a given processor and we consider that old mode tasks $\tau_2$ and $\tau_4$ and the MI task $\tau_1$ are scheduled using EDF on each processor. Processor $\pi_1$ has a total utilization of 93\% both in the old and new modes and processor $\pi_2$ has a total utilization of 100\% in the old mode and 80\% in the new one. Figure~\ref{fig:deadline} presents the EDF schedule on each processor assuming that a mode change request arises at time 7. Up arrows represent task releases and down arrows represent deadlines. In Figure~\ref{fig:deadline}, only releases and deadlines of MD tasks are depicted. The two old mode tasks $\tau_2$ and $\tau_4$ complete their last jobs by time 11. So, at time 11 no more old mode job is pending. The transition delay to switch in the new mode is equal to 4 units of time since the MCR. According to the synchronous mode change protocol, the new mode task $\tau_5$ is enabled at time instant 11 (and simultaneously released in our exemple). The relative transition deadline of $\tau_5$ is $\TD_5=11$ and thus the first job of $\tau_5$ must be completed by time 18. As depicted in Figure~\ref{fig:deadline}, all deadlines are  met, including the transition deadline of $\tau_5$ since its first job is completed by time 15.

\begin{table}
\begin{center}
\begin{tabular}{|c | c | c | c | c | c | c | c|} 
Tasks & Types & Mode & $C_i$ & $T_i$ & $U_i$ & $\pi_i$ & $\TD_i$  \\  \hline
$\tau_1$ & MI & all &1 & 3  & 0.33 & $\pi_1$ & - \\
$\tau_2$ & MD & old &3 & 5  & 0.80 & $\pi_1$ & 20\\
$\tau_3$ & MI & all &4 & 5  & 0.80 & $\pi_2$ & -\\
$\tau_4$ & MD & old &1 & 5 & 0.20 & $\pi_2$ & 20\\ 
$\tau_5$ & MD & new   &3  & 5  & 0.60 & $\pi_1$ & 11\\ \hline
\end{tabular}
\end{center}
\caption{\label{tab:mode} Task Set with two modes}
\end{table}

The delay between a MCR and the current (last) MD job completion defines the \emph{transition latency delay}, denoted $L$. Every MD task $\tau_i$ in the new mode must meet its transition mode deadline $\TD_i$.

\begin{Def} \label{def:td} A mode change with a transition delay $L$ is valid if, and only if, every MD task $\tau_i$ in the new mode satisfies: $L+T_i \leq \TD_i$. 
\end{Def}

Notice that if a mode transition graph can lead to several possible transition changes as previously illustrated in Figure~\ref{fig:graph}. As a consequence, the transition latency delay $L$ to consider for checking transition deadlines according to Definition~\ref{def:td} depends on the current running mode. Let us consider the Figure~\ref{fig:graph} to exhibit two possible cases for determining a worst-case transition delay upper bound. In the one hand, Mode~4 is the only predecessor of Mode~1, thus the worst-case transition delay for checking transition deadlines (i.e., Definition~\ref{def:td}) is $L_4$. In the other hand, Mode~4 has two possible predecessor modes: Mode~3 and Mode~2. At design stage the worst-case transition delay must be considered for checking transition deadlines. Hence, the worst-case transition delay for verifying the transition deadlines of MD task running in Mode~4 is defined by $L=\max(L_3,L_4)$. 

\SubSection{Transition latency delay upper bounds}\label{sec:delay}

Obviously, the transition latency delay depends on the time instant at which the MCR arises. Proving that new mode tasks meet their transition deadlines $\TD_\ell$. Thus, since MCR occurrences are unpredictable, we need to compute a transition latency delay \emph{upper bound}. The assumptions that MI tasks are statically allocated and that none of the MD tasks belongs to several modes ensure that the transition delay in a given mode only depends of the MD tasks executed in that mode.

\begin{Prop} \label{prop:delay} The transition latency delay $L$ only depends on the tasks executed in the current mode, and as a consequence, is independent of MD tasks that will be started in a subsequent mode.
\end{Prop}

Since task to processor allocations meet all timing requirements, then tasks meet all their timing requirements. So, all the MD jobs will be completed by its deadline on each processor. Let $M_i$ be the set of MD task assigned to $\pi_i$ in the current mode, a simple upper bound of the transition delay on processor $\pi_i$ is:
\begin{eqnarray} \label{eq:ub1}
\UB^1_i= \max_{\tau_\ell \in M_i} (T_\ell)
\end{eqnarray}

The second upper bound can be defined by computing a worst-case busy period after a MCR. Let $I_i$ be the set of MI tasks allocated to $\pi_i$, the transition latency delay $L_i$ on processors $\pi_i$ can be defined as follows:
\begin{itemize}
\item the completion of every active MD job.
\item the interference of MI tasks until all MD jobs are completed.
\end{itemize}
Since we consider partitioned scheduling, we can consider each processor separately and we know that in the worst-case, a MCR can arise when all MD tasks have been released and none of them has been scheduled. This delay is bounded by the length of the synchronous busy period of tasks allocated to $\pi_i$ (i.e., initiated by a critical instant) in which one job of each MD task has to be executed and the interference due to MI tasks. This corresponds to the smallest solution $L_i$ of the following fixed-point equation:
\begin{eqnarray}\label{eq:delay}
\forall \pi_i:: L_i=\sum_{\tau_\ell \in M_i} C_\ell  + \sum_{\tau_j \in I_i} \left\lceil \frac{L_i}{T_j} \right \rceil C_j
\end{eqnarray}
The transition latency delay upon processor $\pi_i$ in a given mode is bounded by: 
\begin{eqnarray}\label{eq:ub2}
\UB^2_i=(L_i)
\end{eqnarray}

It is simple to see that upper bounds 1 and 2 are incomparable: if relative deadlines are large numbers in comparison with execution requirements for every MD tasks, then the synchronous busy period will be shorter than all deadlines and hence $\UB^1_i>\UB^2_i$. Conversely, if they are small numbers, then the busy period will be longer than relative deadlines and as a consequence: $\UB^1_i<\UB^2_i$. As a consequence, a transition delay upper bound on the multiprocessor platform is defined by considering the two previously proposed upper bounds for every processor $\pi_i, 1\leq i \leq m$:

\begin{eqnarray}\label{eq:ub}
L=\max_{i=1,\ldots,m} \left(\min(\UB^1_i,\UB^2_i)\right)
\end{eqnarray}

\SubSection{Optimization problems}

The objective is to allocate mode dependent tasks so that the transition latency delays upper bound $L$ is minimized (i.e., Eq.~\ref{eq:ub}). As shown in the previous section, the transition latency delay in a given mode only depends on the tasks executed in the current mode and thus, such an allocation is computed for each system mode. 

We next present two different methods: 
\begin{itemize}
\item offline and optimal method based on a Mixed-Integer Linear Program (MILP) for defining the offline MD task allocations in order to minimize the mode change latency delay and meet all timing constraints.
\item online method in which MD task allocations are performed by an online algorithm and we provide a sufficient schedulability test that can be checked offline (i.e., at the design stage of the system).
\end{itemize}

%------------------------------------------------------------------------- 
\Section{Exact offline allocation method}

Allocating tasks to processors in order to minimize an objective function is a combinatorial problem known to be $\mathcal{NP}$-hard. Mathematical programming is a common way to model and solve optimally such problems. In the real-time scheduling literature, mixed linear programming has been investigated for checking feasibility of uniprocessor real-time scheduling problems~\cite{Seto98, Bini:09,Zeng:10} and for optimally allocating real-time tasks~\cite{Bini:08}. To the best of our knowledge, mathematical programming has never been studied for analyzing real-time systems subjected to mode changes.

We shall define a MILP for each mode that computes the MD task allocation while minimizing the transition latency delay $L$ from the considered mode. Transition deadlines are then checked using the optimal value $L$ computed by the MILP (i.e., transition timing constraints in Definition~\ref{def:td}).

The proposed MILP tackles with two sets of constraints:
\begin{itemize}
\item Constraints ensuring feasible task assignment: every MD task is assigned to a processor and the utilization of every processor is less than or equal to 1.
\item Constraints computing the transition delay upper bound.
\end{itemize}

We now details these two sets of constraints in separate sections, then the whole MILP will be presented. Table~\ref{tab:notations} summarizes notations that are used hereafter.

\begin{table}
\begin{center}
\begin{tabular}{| c | l |}

Symbols & Comments\\ \hline
$m$ & number of identical processors\\
$M$ & Set of mode dependent tasks\\
$M_i$ & Set of mode dependent tasks upon $\pi_i$\\
$I$ & Set of independent tasks\\
$I_i$ & Set of mode independent tasks upon $\pi_i$\\
$C_i$ & task $\tau_i$ worst-case execution time\\
$T_i$ & task $\tau_i$ period\\
$U_i$ & task $\tau_i$ utilization\\
$y_{i,j}$ & binary variable =1 if $\tau_j$ is allocated upon $\pi_i$\\
$p_i$ & binary variable for disjunctive constraints\\
$x_j$ & number of jobs of $\tau_j$ in the busy period \\
$L_i$& busy period length on $\pi_i$\\
$L$ & real number (Transition delay upper bound)\\
$\HV$ & Arbitrary huge value (disjunctive constraints)\\\hline

\end{tabular}
\caption{\label{tab:notations} Notations in mathematical programs}
\end{center}
\end{table}

\SubSection{Feasible allocation}

Let $y_{i\ell}$ be a binary variable such that $y_{i\ell}=1$ if MD task $\tau_\ell$ ($\ell \in M$) is allocated to $\pi_i$, $0$ otherwise. In order to ensure that every MD task will be allocated upon only one processor, the following constraint must be checked:
\begin{eqnarray}
\begin{array}{l l l}
\sum_{i=1}^m y_{ij} = 1 & & j \in M.\\
\end{array}
\end{eqnarray}

Furthermore, every processor must satisfy that the total utilization of allocated MD and MI tasks is less than or equal to 1:

\begin{eqnarray}
\begin{array}{l l l}
\sum_{\ell\in M} y_{i\ell}U_\ell + \sum_{\ell \in I_i} U_\ell &\leq 1 & i = 1, \ldots, m.\\
\end{array}
\end{eqnarray}

\SubSection{Transition delay upper bounds}

As shown in Equation~\ref{eq:ub}, the upper bound of the transition delay is computed as the minimal value between two distinct upper bounds. We first present them separately, and then we shall present how to extend these constraints in order to model the choice of the minimal value among $\UB^1_i$ and $\UB^1_i$ on every processor $\pi_i$.

Let $L \in \mathbb{R}$ be the optimized value of the transition delay upper bound. Since $\UB^1_i=\max_{\ell\in M_i}(T_\ell)$ Eq.~\ref{eq:ub1}, (minimizing $L$ such that $L \geq T_\ell, \forall \ell \in M_i$) directly computes $\UB^1_i$. This leads to the following first set of constraints:
\begin{eqnarray}
y_{ij} T_j \leq L \qquad i = 1, \ldots, m; j \in M
\end{eqnarray}
 
We now consider $\UB^2_i$ defined in Equation~\ref{eq:ub2}. Due to ceiling functions, $\UB^2_i$ is not linear and furthermore corresponds to the maximum of the smallest solutions of a fixed-point equations $(L_i)$. The linearization of $L_i$ is based on the techniques introduced by~\cite{Seto98}. Let $x_j \in \mathbb{N}$, modeling the number of jobs of a MI task $\tau_j$ interfering in the transition delay. $\left\lceil \frac{L_i}{T_j} \right \rceil C_j$ corresponds to the worst-case interference in interval of time $[0,L_i)$. Since $L_i$ will be minimized for each processor (i.e., the busy period length), then $x_j = \left\lceil \frac{L_i}{T_j} \right\rceil$. Thus, the longest busy period then corresponds to the complete execution of MD jobs that is equal to $\sum_{\ell\in M} y_{i\ell}C_\ell$, plus the interference of MI tasks: $\sum_{\ell\in I_i} x_\ell C_\ell$. Such a busy period will be completed by time $x_jT_j$ (i.e., before the release of subsequent jobs) and leads to an additional constraint in the mathematical program. 

For convenience, we define $L_i$  as the busy period length on processor $\pi_i$. $L_i$ does not correspond to a variable in the mathematical program, but every time $L_i$ will be used it will be replaced by the right-hand side of the following equation: 
\begin{eqnarray}\label{eq:Li}
L_i=\sum_{\ell \in M} y_{i\ell}C_\ell + \sum_{\ell \in I_i} x_\ell C_\ell & i = 1, \ldots, m
\end{eqnarray}
Hence, the constraint checking the end of the longest busy period can be stated as: $L_i  \leq x_jT_j \;i = 1, \ldots, m; j \in I_i$.

The last problem is now to compute for every processor $\pi_i, 1\leq i \leq m$ which upper bound among $\UB_1$ and $\UB_2$ will lead to the smallest value of $L$. 
The corresponding constraints are disjunctive: $ L_i\geq \UB_1 \textbf{ or } L_i\geq \UB_2$. Such disjunctive constraints can be represented by linear constraints with a binary variable. Let $p_i \in \{0,1\}$ and $\HV$ be an arbitrary high value, then the previous disjunctive constraint is modeled by the two linear constraints:
\begin{alignat}{4}
   \UB^1_i &\leq L &&+ &(1-p_i)\HV&&\\
   \UB^2_i &\leq L &&+ &p_i \HV&&
\end{alignat}

The disjunction of the two previous constraints is obtained as follows: if $p_i=1$ then the second constraint is always satisfied and thus only the first one is taken into account while computing the optimal solution; if $p_i=0$, it is the reverse situation and thus the second constraint is effective during the optimization process. Now, replacing $\UB^1_i$ and $\UB^2_i$ by their corresponding values, the final set of constraints is finally obtained:

\begin{eqnarray*}
\begin{array}{l l}
L_i=\sum_{\ell \in M} y_{i\ell}C_\ell + \sum_{\ell \in I_i} x_\ell C_\ell   & i = 1, \ldots, m.\\
L_i \leq x_jT_j & i = 1, \ldots, m; j \in I_i.\\
L_i \leq L + (1-p_i)\HV & i = 1, \ldots, m.\\
y_{ij} T_j \leq L + p_i \HV & i = 1, \ldots, m; j \in M.\\
y_{ij}\in \{0,1\}, p_i\in \{0,1\}; & i = 1, \ldots, m; j \in M.\\
L\in \mathbb{R}, L_i \in \mathbb{R}; x_\ell \in \mathbb{N} &i = 1, \ldots, m; \ell \in I\\
\end{array}
\end{eqnarray*}

\SubSection{MILP formulation}

Constraints defined in previous sections are now merged to define the complete MILP formulation in Figure~\ref{fig:milp}. We remove $L_i$'s by their definitions everywhere that have been used. Let $n$ be the number of tasks in the considered mode: $n=\left\vert{I}\right\vert + \left\vert{M}\right\vert$, the MILP has: 
\begin{itemize}
\item $m(n+2)$ constraints,
\item $m(\left\vert{M}\right\vert+1)$ binary variables,
\item $\left\vert{M}\right\vert$ integer variables,
\item 1 real variable.
\end{itemize}

Hence, the number of variables and constraints are polynomially bounded in the size of the input problem (i.e.\@ platform and task set sizes for the considered mode). The MILP is solved for every existing mode at the design stage. The corresponding allocation tables are then stored in the system memory and used at running time for allocating MD tasks of the new mode whenever a MCR transition phase is completed.

\begin{figure*}
\begin{center}
\begin{eqnarray}
\begin{array}{l l l l}
\text{Minimize}&  L  \label{milp1:opt}\\
\text{subjected to} & \nonumber \\
 &\sum_{i=1}^m y_{ij} = 1 & & j \in M \label{milp1:1}\\
&\sum_{\ell\in M} y_{i\ell}U_\ell + \sum_{\ell\in I_i} U_\ell \leq 1 & & i = 1, \ldots, m \label{milp1:2}\\
&\sum_{\ell\in M} y_{i\ell}C_\ell + \sum_{\ell\in I_i} x_\ell C_\ell  \leq x_jT_j & & i = 1, \ldots, m; j \in I_i \label{milp1:3}\\
&\sum_{\ell\in M} y_{i\ell}C_\ell + \sum_{\ell\in I_i} x_\ell C_\ell  \leq L + (1-p_i)\HV & &i = 1, \ldots, m \label{milp1:4}\\
&y_{ij} T_j \leq L + p_i\HV & & i = 1, \ldots, m; j \in M \label{milp1:5}\\
&y_{ij}\in \{0,1\}   & & i = 1, \ldots, m; j \in M \nonumber\\
&p_i\in \{0,1\}  & & i = 1, \ldots, m \nonumber\\
& x_\ell \in \mathbb{N}  & & \ell \in I \nonumber\\
&L\in \mathbb{R} \nonumber\\
\end{array}
\end{eqnarray}
\end{center}
\caption{MILP formulation of the offline allocation method \label{fig:milp}}
\end{figure*}

\SubSection{Case study}

\begin{figure}
\begin{center}
\begin{tikzpicture}[->,>=stealth',shorten >=1pt,auto,node distance=3cm,
  thick,main node/.style={circle,fill=blue!20,draw,font=\sffamily\Large\bfseries}]

  \node[main node] (1) {1};
  \node[main node] (2) [right of=1] {2};

  \path[every node/.style={font=\sffamily\small}]
    (1) edge [bend right] node[below] {$L_1$} (2)
    (2) edge [bend right] node[above] {$L_2$} (1);
\end{tikzpicture}
\end{center}
\caption{\label{fig:modes} Case Study: possible transitions between modes}
\end{figure}
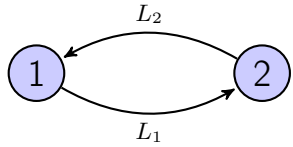

In order to illustrate the problem and the proposed methods, we shall consider a simple case study. The platform is composed of two-identical processors. The system is based on two distinct running modes. The system is started in the first mode and then alternates indefinitely between the two modes as depicted in Figure~\ref{fig:modes}. The task set is defined by: 4 mode-independent (MI) tasks $I=\{\tau_1,\tau_2,\tau_3,\tau_4\}$ and 6 mode-dependent (MD) tasks with in Mode~1: $D_1=\{\tau_5,\tau_6,\tau_7,\tau_8,\tau_9\}$ and in Mode~2: $D_2=\{\tau_{10}\}$. The allocation of mode independent tasks is static and known a priori. The total utilization of MI tasks are respectively $0.66$ for $\pi_1$ and $0.36$ for $\pi_2$. Tasks parameters are summarized in Table~\ref{tab:taskset}.

\begin{table}
\begin{center}
\begin{tabular}{|c | c | c | c | c | l | c | c|} 
Tasks & Types & Mode & $C_i$ & $T_i$ & $U_i$ & $\pi_i$ & $\TD_i$  \\  \hline
$\tau_1$ & MI & all &10 & 30  & 0.333 & $\pi_1$ & - \\
$\tau_2$ & MI & all &20 & 60  & 0.333 & $\pi_1$ & -\\
$\tau_3$ & MI & all &15 & 90  & 0.16 & $\pi_2$ & -\\
$\tau_4$ & MI & all &20 & 100 & 0.20 & $\pi_2$ & -\\ \hline 
$\tau_5$ & MD & 1   &7  & 40  & 0.175 & $-$ & 150\\
$\tau_6$ & MD & 1   &1  & 10  & 0.10 & $-$ & 100\\
$\tau_7$ & MD & 1   &1  & 20  & 0.05 & $-$ & 150\\
$\tau_8$ & MD & 1   &2  & 30  & 0.0666 & $-$ & 200\\
$\tau_9$ & MD & 1   &3  & 25  & 0.12 & $-$ & 200\\\hline
$\tau_{10}$ & MD &2 &50 & 100 & 0.50 & $-$ & 150\\\hline
\end{tabular}
\end{center}
\caption{\label{tab:taskset} Case Study: Task characteristics with two modes}
\end{table}

The mathematical program defined in Figure~\ref{fig:milp} has been solved for the task set presented in Table~\ref{tab:taskset} in order to compute the optimal allocation of MD tasks for the two modes. We used Frontline solver~\cite{solver} to solve MILP models.

\subsubsection{Mode 1 and transition delay}

For the first mode, the MILP leads to the following MD task allocations:
\begin{itemize}
\item upon $\pi_1$, the allocated MD tasks are:  $M_1=\{\tau_5,\tau_6\}$, its total utilization with MI tasks is $0.941666\ldots$.
\item upon $\pi_2$, the allocated MD tasks are: $M_2=\{\tau_7,\tau_8,\tau_9\}$, its total utilization with MI tasks is $0.60333\ldots$.
\end{itemize}

The computed optimal transition delay upper bound is $L=40$, corresponding to the saturated constraint on $\UB^1_2$, that is the maximal period of the MD tasks allocated upon $\pi_2$ (i.e., $\tau_5$). The constraint related to $\UB^1_1$ is saturated for task $\tau_9$ which has the greatest period among tasks allocated to $\pi_1$. Table~\ref{tab:bound} gives the values of the transition latency delay upper bound for every processor.  In this case study, since MD task periods are small, $\UB^1_i$ dominates $\UB^2_i$ for both processors $\pi_i, 1 \leq i \leq 2$. As a consequence, the values related to $\UB^2_i$ presented in Table~\ref{tab:bound} have been computed using fixed point equations since the corresponding constraints are not saturated in the final program due to disjunctive constraints (i.e., $p_1=p_2=0$ in the optimal program).

\begin{table}
\begin{center}
\begin{tabular}{|c ||c |c ||c |c |}
\hline
Modes /&  \multicolumn{2}{c||}{Mode 1} & \multicolumn{2}{c|}{Mode 2} \\
Processors & $\UB^1_i$ & $\UB^2_i$ & $\UB^1_i$ & $\UB^2_i$ \\ \hline
$\pi_1$ & 25 & 48 & -- & -- \\ 
$\pi_2$ & 40 & 41 & 100 & 85 \\ \hline
\end{tabular}
\caption{Computed transition delay upper bounds for mapping MD tasks in Modes 1 and 2\label{tab:bound}}
\end{center}
\end{table}

According to the worst-case transition delay computed by the MILP, $\tau_{10}$'s transition deadline (i.e., $\TD_{10}=150$) must be checked. The MILP solution guarantees that the EDF will schedule all the tasks in the new mode. Thus, the sufficient condition to check this transition deadline is satisfied: $L+T_{10} = 40 + 100 \leq \TD_{10}$.

\subsubsection{Mode 2 and transition delay}

According to total utilizations of MI task on each processor, the unique MD task (namely $\tau_{10})$ has an utilization of 0.5 and thus can only be allocated on $\pi_2$. Hence, the MILP solver allocates $\tau_{10}$ to $\pi_2$ leading to the transition delay upper bound $L=85$. This value corresponds to the second bound $\UB_2^2$: the length of the synchronous busy period on $\pi_2$ (i.e., the interference of the MI tasks $\tau_3$ and $\tau_4$ during the completion of $\tau_{10}$). In this case, the upper bound corresponding to the maximum of MD task periods $\UB_1^2$ is dominated by the upper bound based on the busy period length. The corresponding values are presented in Table~\ref{tab:bound}.

To complete the case study in the offline setting, MD task transition deadlines must be checked. Using the transition delay $L=85$, one can easily check that the transition deadlines of $\{\tau_5,\tau_6,\tau_7,\tau_8,\tau_9\}$ are all satisfied.

%------------------------------------------------------------------------- 
\Section{Online allocation method}\label{sec:online}

Linear programs can be solved in polynomial time but introducing binary or integer variables in the math programming model leads to a strongly $\mathcal{NP}$-hard problem. Thus, solving such large-scaled optimization problems is quite difficult in practice and the MILP solver is not able to converge to integral values of decision variables nor to establish optimality conditions. Another drawback of the offline approach is that task allocation tables corresponding the different system modes must be stored in the memory of the real-time system.

Two main approaches can be used to cope with large-scaled optimization problems:

\begin{itemize}
\item defining an offline heuristics (with a reasonable time complexity losing optimality) for solving the original allocation problem such as simulated annealing or genetic algorithms. As in the exact method, a static allocation table stores the mapping of all tasks.
\item using online allocation algorithms such as well-known online bin packing heuristics~\cite{Gal95} and providing conditions ensuring feasibility at runtime. In this case whenever a MCR occurs, the tasks to be started in the new mode are allocated with an online algorithm that is executed when the last task running in the old mode completes.
\end{itemize}
The latter approach is investigated hereafter assuming that (i) EDF is locally used on every processor to schedule the tasks and (ii) MD task allocations is performed by the First-Fit Decreasing algorithm~\cite{Lopez:04}. First the First-Fit Decreasing algorithm sorts the MD tasks into non-increasing order of their utilization, and then processes them in that order by allocating each MD task on the first processor into which it fits without exceeding its capacity. As in the offline setting, we assume that MI tasks are statically allocated and are not allowed to migrate at runtime. We make no particular assumption about the algorithm used to allocate MI tasks. 

Two problems need to be solve for using online allocation after a MCR~:
\begin{itemize}
\item the computed allocation must be feasible. We shall use the results presented in~\cite{Lopez:04} for checking that allocations will lead to feasible schedule on each processor.
\item the mode transition delay must be bounded in order to ensure that transition deadlines will be met for every MD task. We will define an algorithm that will solve several mathematical programs (i.e., one for each processor) for computing a transition delay \emph{upper bound} of the platform.
\end{itemize}

We detail the solution to these problems in the two next sections.

\SubSection{Feasible allocation}\label{sec:falloc}

A \emph{reasonable allocation algorithm} is one which fails to allocate a task only when there is no processor in the system which can hold it. MI tasks are statically allocated by designers and MD tasks are allocated by the First-Fit Decreasing. So, such an allocation of the tasks can always correspond to a First-Fit allocation. Since tasks can be renumbered so that the considered allocation correspond to a First-Fit allocation. As a consequence, the overall allocation process (MI and MD tasks) is reasonable. This property will allow to use known feasibility results on EDF scheduling of partitioned multiprocessor real-time systems.

Utilization bounds of bin packing algorithms have been studied in~\cite{Lopez:04} for real-time sporadic tasks with implicit-deadlines. Precisely, let $\beta$ be the maximum number of tasks which can fit into one processor under EDF scheduling:
\begin{eqnarray}
\beta=\left\lfloor \frac{1}{U_{\max}} \right\rfloor
\end{eqnarray}
The maximum utilization bound of any reasonable algorithm, First-Fit and Best-Fit is~\cite{Lopez:04} is defined:
\begin{eqnarray} \label{eq:onlinetest}
U_{\text{sum}} \leq \frac{\beta m+1 }{\beta+1}
\end{eqnarray}

Such a maximum utilization bound is tight for reasonable algorithms including First-Fit and Best-Fit~\cite{Lopez:04}. So, there exists task set having a total utilization just beyond that bound cannot be partitioned over the processors. But, any task set having a total utilization less than or equal to this tight utilization bound  is guaranteed to be feasible under EDF scheduling.

The sufficient schedulability test stated in Inequality~(\ref{eq:onlinetest}) will be used to check whether a processor allocation will lead to a feasible EDF schedule or not.

\SubSection{Transition latency upper bound}

The second concerns is to compute the upper bound of any mode transition delay. Since online allocations are used whenever a MCR occurs, it is not possible to predict which allocation will define the worst-case transition delay at runtime. As a consequence, we need to define an upper bound of the mode transition delay while considering all possible MD task allocations. For that will first compute the worst-case allocation of MD tasks for every processor $\pi_i, 1 \leq i \leq m$. Let $M_i$ the subset of MD tasks allocated on processor $\pi_i$ using such a principle, we then compute the worst-case delay $L_i$ using Equation~\ref{eq:delay}. The upper bound of the transition delay will be the longest computed delay among all $L_i, 1 \leq i \leq m$. We will see that choosing which MD tasks have to be considered is equivalent to solve a knapsack problem.

As shown previously, the transition delay only depends on the MD tasks currently running before a MCR. It follows from Equation~\ref{eq:delay} that $L_i$ is non-decreasing, and hence will be maximized when the cumulative length of MD tasks allocated to $\pi_i$ is as large as possible. For every processor $\pi_i$, we have to compute which subset of $M_i \subseteq M$ maximizes the length of selected tasks: $z_i=\sum_{\ell \in M_i} C_{\ell}$. Before detailing how to compute $z_i$ using mathematical programming, we summarize the overall approach for a given mode in Algorithm~\ref{alg:upperbound}.

\begin{algorithm*}[t]
\SetKwInOut{Input}{input}
\SetKwInOut{Output}{output}
\caption{\label{alg:upperbound}Transition Latency Delay Upper Bound}
  \SetLine
  \Input{\\$\{C_i,T_i\}, 1\leq i \leq n $ : array of real \tcc*[f]{Task parameters}\\
  $I_i, 1\leq i \leq m$ : set of MI tasks \tcc*[f]{Allocated MI tasks upon $\pi_i$}\\
   $M$ : set of MD tasks \tcc*[f]{Set of MD tasks}\\
 }
  \Output{$L$ : real \tcc*[f]{The transition latency delay upper bound}\\}
  \BlankLine
  $L:=0$\;
  \ForEach{$i:=1,\ldots, m$}{
        $z_i:=Solve(I_i,M);$ \tcc*[h]{Compute the longest possible transition delay upon $\pi_i$}\\
        $L'_i:=z_i$ \;
        \Repeat(\tcc*[f]{Solve fixed-point equation}){$L_i=L'_i$}{
            $L_i:=L'_i$\;
            $L'_i:=z_i+\sum_{\ell \in I_i} \left\lceil \frac{L_i}{T_{\ell}} \right\rceil C_{\ell}$\;       
        }(\tcc*[f]{smallest fixed-point is reached})
  }
  $L:= \max_{i=1\cdots m} (L_i)$\;
  \Return{$L$}\;
\end{algorithm*}

Computing a subset of $M_i \subseteq M$ that maximizes the length of selected old jobs is equivalent to solve 0--1 knapsack problem. We recall the problem definition: given a knapsack with maximum capacity $W$, and a set $S$ consisting of $n$ items, each item $i$ has some weight $w_i$ and benefit value $b_i$ (all $w_i$ , $b_i$ and $W$ are positive numbers), the problem to solve is: how to pack the knapsack to achieve maximum total value of packed items? This latter problem is known to be $\mathcal{NP}$-hard in the weak sense and can be solve in pseudo-polynomial time with dynamic programming~\cite{Martello:90}. So, large-scaled instances can be efficiently solved. Numerous techniques have been proposed for solving efficiently this combinatorial problem. Since the problem will be solve offline, we focus on exact methods.

We describe a mathematical formulation of selecting a subset of MD tasks leading to the longest transition delay on a given processor $\pi_i$ subjected that such the computed allocation leads to a feasible schedule (i.e., the processor utilization is less than or equal to one). In the worst-case, exactly one job of the MD task of the old mode must be completed. Thus, the objective is to:
\begin{eqnarray}
z_i=\sum_{\ell \in M} C_{\ell}
\end{eqnarray}

The constraint to enforce the selected MD tasks are feasible with preallocated MI tasks is: 
\begin{eqnarray}
\sum_{\ell \in M} U_\ell+\sum_{\ell \in I_i} U_\ell \leq 1
\end{eqnarray}

We next provide the mathematical program for computing $z_i$: let $y_{\ell}$ be a binary variable equals to 1 if $\tau_{\ell} \in M$ is in the selected subset, 0 otherwise and let $I_i$ be the set of MI tasks statically allocated to $\pi_i$. The available utilization on $\pi_i$ before allocating MD tasks is bounded by $1-\sum_{\ell\in I_i} U_\ell$. Thus, the optimization problem is to maximize the total length of selected job subjected to the constraint that the total utilization of allocated tasks upon $\pi_i$ is not greater than one (i.e., EDF can produce a feasible schedule upon $\pi_i$). The corresponding mathematical program is presented in Figure~\ref{fig:milp2}.

\begin{figure}

\begin{eqnarray}
\text{Maximize}     &  z_{i}=\sum_{\ell \in M} y_{\ell}C_{\ell}  \label{milp2:opt}\nonumber\\
\text{subjected to} &\nonumber\\
                    &\sum_{\ell\in M} y_{\ell}U_\ell \leq 1- \sum_{\ell\in I_i} U_\ell \label{milp2:c1}\\[2ex]
                    & y_{\ell}\in \{0,1\}  \qquad   \ell \in M
\end{eqnarray}

\caption{\label{fig:milp2} 0--1 knapsack formulation of the MD task selection problem on processor $\pi_{i}$} 
\end{figure}

\SubSection{Case study}

We illustrate the principles previously presented on the case study presented in Table~\ref{tab:taskset} for Modes~1 and~2. In both cases, the sufficient feasibility condition of the online allocation process must be established and the transition delay upper bound must be computed to verify if MD task transition deadlines will be met at runtime. 

\subsubsection{Validation of Mode 1}

\paragraph{FF/EDF Feasibility.}

Using feasibility condition presented in Section~\ref{sec:falloc} for the tasks running in the first mode, we have  $U_{\max}=1/3$ and $U_{\text{sum}}=1.545$. Hence, $\beta=3$  and the maximum utilization bound that can be guaranteed using a First-Fit allocation of MD tasks is equal to $\frac{m\beta+1}{\beta+1}=\frac{7}{4}=1.75$. Thus, the proposed allocation algorithm based on First-Fit allocation of MD tasks will always be able to allocate MD tasks if EDF is used on each processor since $U_{\text{sum}}=1.545 \leq 1.75$.

\paragraph{Checking Transition deadlines.}

The MD tasks starting in the Mode~1 will be delayed be the completion of task completed in Mode~2 after a MCR. As shown previously, $\tau_{10}$ is the unique MD task in that mode and can only be allocated on $\pi_2$. Thus, there is no subset selection problem and the unique solution is $z_2=C_{10}=50$. Inserting such a value in the fixed point equation (i.e. Equation~\ref{eq:ub2}) leads to: $L_2=85$. Notice that such computations have already be done while analyzing the processor $\pi_2$ in the Offline Method and we already concluded that transition deadlines of tasks $\{\tau_5,\tau_6,\tau_7,\tau_8,\tau_9\}$ are satisfied while switching from Mode~2 to Mode~1.

\subsubsection{Validation of Mode 2}

\paragraph{FF/EDF Feasibility.}

In Mode~2, the operational task set is defined by $U_{\max}=1/2$ and $U_{\text{sum}}=1.5333\ldots$.  Hence, $\beta=2$ and the maximum utilization allowed by the Lopez's et \emph{al.} sufficient schedulability condition is $\frac{m\beta+1}{\beta+1}=\frac{5}{3}= 1.666\ldots$. As a consequence, the proposed online allocation process of MD tasks will lead to feasible schedules in Mode~2 upon First-Fit allocation of MD tasks and the partitioned EDF scheduling policy since the total utilization of tasks in Mode~2 is $1.5333\ldots$ (i.e. the maximum guaranteed utilization bound).

\paragraph{Checking Transition deadlines.} 

The transition delay upper bound while launching MD tasks in Mode~2 is only due to the previous running mode before starting this new mode. In the case study, the predecessor of Mode~2 is the Mode~1. Hence, the transition delay upper bound will be computed as follow according to the tasks to stop in the first mode:
\begin{itemize}
\item Processor $\pi_1$: two MI tasks are preallocated (i.e. $\tau_1$ and $\tau_2$) leading to a cumulative utilization of $0.666\ldots$. Thus, the remaining utilization for MD tasks is $0.333\ldots$. Selecting MD tasks using the MILP presented in Figure~\ref{fig:milp2} leads to select $\tau_5$ and $\tau_9$ and the corresponding value of the objective function is $z_1=10$. Introducting that value while solving the fixed-point equation leads to $L_1=50$. 
\item Processor $\pi_2$: $\tau_3$ and $\tau_4$ are now preallocated and have a cumulative utilization equal to 0.36. Thus, the maximum possible cumulative utilization for MD tasks upon $\pi_2$ is bounded by 0.63. The solver selects all MD tasks and the objective function is maximized for $z_2=14$. Thus, solving the fixed-point equation leads to $L_2=49$.
\end{itemize}

Hence, $L=\max(L_1,L_2)=50$. This worst-case delay can be incurred whenever a mode change from Mode~1 to Mode~2 occurs. Thus, only task $\tau_{10}$ is concerned by this transition delay before starting the Mode~2. The transition deadline of $\tau_{10}$ is 150 and the sufficient condition $L+T_{10} = 50+100 \leq \TD_{10}$ is satisfied.

%------------------------------------------------------------------------- 
\Section{Conclusion}
%------------------------------------------------------------------------- 

The paper presented two methods for handling mode changes in partitioned scheduling upon multiprocessor platform. Tasks are scheduled using the EDF scheduling policy on every processor. We consider two kind of implicit-deadline tasks: mode independent tasks that are executed in all modes and mode dependent tasks that are run in a single mode. The considered mode change protocol is synchronous meaning that it stops any mode dependent tasks before launching the tasks in the new mode. This behavior ensures a temporal isolation of mode dependent tasks running in different modes. The first method is offline and computes a static allocation of MD tasks upon each processor and for each mode. The transition delay upper bounds for switch from the one mode to the subsequent one have been proposed. A transition delay is incurred by the new mode dependent task that must meet their transition deadlines. In the second approach we provide sufficient conditions for verifying that online First-Fit based allocation of mode dependent tasks will leads to feasible schedules and allow to satisfy task transition deadlines. Both approaches have been completely illustrated on a simple case study.

In future work, we want to extend the methods for allowing Mode Independent task to migrate when a mode change is occurring. This should allow a higher utilization of the platform and also to manage the workload balancing upon processors. Another perspective is to consider tasks with constrained and arbitrary deadlines.

\bibliographystyle{acm}
\bibliography{latex8}

\end{document}